\documentclass[aps,pre,reprint,amsmath,amsfonts,amssymb,showkeys,superscriptaddress,floatfix]{revtex4-2}
\usepackage[utf8]{inputenc}
\usepackage{graphicx}
\usepackage{xspace}
\newcommand{\cretin}{\textsc{Cretin}\xspace}
\newcommand{\jjatom}{\textsc{jjatom}\xspace}

\newcommand{\supp}{Supporting Information\xspace}
\begin{document}
\title{Decreasing ultrafast X-ray pulse durations with saturable absorption and resonant transitions}
\author{Sebastian Cardoch}
\email{sebastian.cardoch@physics.uu.se}
\affiliation{Department of Physics and Astronomy, Uppsala University, Box 516, SE-751 20, Uppsala, Sweden}
\author{Fabian Trost}
\affiliation{Center for Free-Electron Laser Science, Deutsches-Elektronen Synchrotron (DESY), Hamburg, Germany}
\author{Howard A. Scott}
\affiliation{Lawrence Livermore National Laboratory, L-18, P.O. Box 808, 94550, Livermore, CA, USA}
\author{Henry N. Chapman}
\affiliation{Center for Free-Electron Laser Science, Deutsches-Elektronen Synchrotron (DESY), Hamburg, Germany}
\affiliation{The Hamburg Center for Ultrafast Imaging, Universität Hamburg, Luruper Chaussee 149, 22761 Hamburg, Germany}
\affiliation{Department of Physics, Universität Hamburg, Luruper Chaussee 149, 22761, Hamburg, Germany}
\author{Carl Caleman}
\affiliation{Department of Physics and Astronomy, Uppsala University, Box 516, SE-751 20, Uppsala, Sweden}
\affiliation{Center for Free-Electron Laser Science, Deutsches-Elektronen Synchrotron (DESY), Hamburg, Germany}
\author{Nicusor Timneanu}
\email{nicusor.timneanu@physics.uu.se}
\affiliation{Department of Physics and Astronomy, Uppsala University, Box 516, SE-751 20, Uppsala, Sweden}
\date{\today}
\begin{abstract}
Saturable absorption is a nonlinear effect where a material's ability to absorb light is frustrated due to a high influx of photons and the creation of electron vacancies. Experimentally induced saturable absorption in copper revealed a reduction in the temporal duration of transmitted X-ray laser pulses, but a complete understanding of this process is still missing. In this computational work, we employ non-local thermodynamic equilibrium plasma simulations to study the interaction of femtosecond X-rays and copper. Following the onset of frustrated absorption, we find that a \(K\mbox{--}M\) resonant transition occurring at highly charged states turns copper opaque again. The changes in absorption generate a transient transparent window responsible for the shortened transmission signal. We also propose using fluorescence induced by the incident beam as an alternative source to achieve shorter X-ray pulses. Intense femtosecond X-ray pulses are valuable to probe the structure and dynamics of biological samples or to reach extreme states of matter. Shortened pulses could be relevant for emerging imaging techniques.
\end{abstract}
\keywords{X-ray, copper, saturable absorption, frustrated absorption, temporal shape, femtosecond pulse, NLTE theory, K-shell fluorescence, incoherent diffractive imaging, warm dense matter, free-electron lasers}
\maketitle
\section{Introduction}
X-ray free-electron lasers (XFELs) can generate pulses with unprecedented characteristics suitable to study the structure and dynamics of biological samples~\cite{mancuso_single_2019}, ultrafast phase transitions~\cite{makita_femtosecond_2019}, or exotic states of matter~\cite{lee_finite_2003}. A current goal is to produce high-intensity (\(10^{17}\mbox{--}10^{19}\)~W/cm\(^{2}\)) extremely short pulses of tens of femtosecond that can image matter at Ångström resolution before the onset of radiation damage or atomic motion~\cite{neutze_potential_2000,chapman_femtosecond_2011}. Recent suggestions for a new technique, incoherent diffractive imaging~\cite{classen_incoherent_2017}, require the development of X-ray pulses shorter than the coherence time of fluorescence emission~\cite{trost_photon_2020}. The intense pulses from XFELs can alter the structure and optical properties of materials, resulting in nonlinear effects. Taking advantage of this material response, \citet{inoue_shortening_2021} experimentally demonstrated temporal shortening of X-rays by inducing saturable absorption in a solid copper target, thus uncovering a potential approach to satisfy the pulse constraints for incoherent imaging.

Saturable absorption, which describes fluence-induced transparency, has been investigated in the soft and hard X-ray regimes on transitions metals such as aluminum~\cite{nagler_turning_2009} and iron~\cite{yoneda_saturable_2014}. The initially opaque target attenuates the incoming radiation until depletion of electrons in the K-shell weakens Coulomb interactions with the core, causing broadening and shifting of the K-edge to higher energies~\cite{inoue_shortening_2021}. The sample achieves this transparent state if the photoionization rate is comparable to the Auger-Meitner and fluorescence decay rates~\cite{young_femtosecond_2010}. \citet{inoue_shortening_2021} indirectly measured the transmission of X-rays through the material and found a detectable temporal decrease compared to the incident beam at a few selected fluences. The study opened interesting questions about the dynamic processes inside the material. With a greater photon flux, we expect an increased formation of single core-hole states, a faster shift in copper's K-edge, and the material will reach transparency sooner. If the time it takes to go from cold absorption to saturation exclusively dictates the transmission of X-rays, the resulting pulse duration should increase at higher fluence, contradicting experimental evidence. We identify a more complete description of the electronic damage that governs transmission is needed.

In this paper, we computationally investigate why XFEL beams transmitted through copper have shorter temporal durations. We also explore Cu fluorescence, induced by absorption of the incident beam, as an alternative source of X-rays that might exhibit similar temporal characteristics. We chose a copper target to compare our calculations with the results of the experiment performed by \citet{inoue_shortening_2021}. Copper has a fluorescence yield comparable to Auger-Meitner electron yield with its K\({\alpha}\) emission found above iron's, cobalt's, and nickel's K-edge. Transmission or fluorescence originating from the copper target can generate core vacancies on these lower Z elements, found in crystals or biomolecules, whose fluorescence could be applied for structure determination~\cite{classen_incoherent_2017}.

High-intensity X-rays with wavelengths just above copper's K-edge experience significant absorption in the material (absorption coefficient \(10^{3}~\mathrm{cm^{-1}}\)). Large quantities of energy are deposited mainly from \(1s\) electron ionization leading to further damage to the electronic structure, and the sample becomes a plasma within femtoseconds after exposure~\cite{chapman_femtosecond_2006}. Photon-matter collisions create a cascade of secondary processes and a dynamic radiation energy landscape that results in notable temperature differences between ions and electrons and between the front (facing the beam) and back of the sample~\cite{gamaly_physics_2011}. Thermalization and cooling through expansion occur on much longer timescales (\(1\mbox{--}10\)~ps), so the material exists in a transient warm-dense-matter state that can be studied by non-local thermodynamic equilibrium (NLTE) theory~\cite{hummer_formation_1971,chung_applications_2009,chung_extension_2007,rosen_role_2011,hau-riege_introduction_2011}.

\begin{figure}
    \centering
    \includegraphics[width=\linewidth]{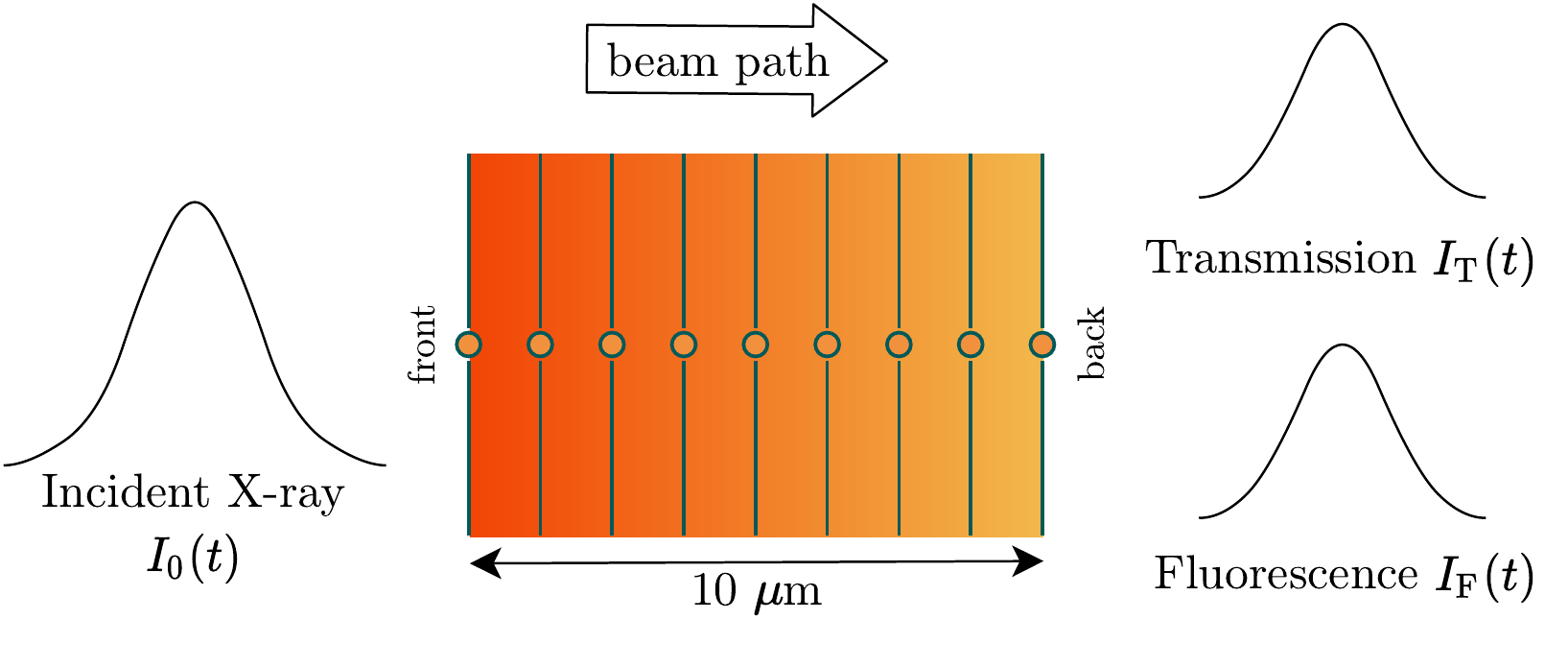}
    \caption{\label{fig:diagram}Schematic representation of a 1D simulation that follows the effects of an incident Gaussian pulse \(I_\mathrm{0}(t)\) in the material and monitors transmission \(I_\mathrm{T}(t)\) and fluorescence \(I_\mathrm{F}(t)\) intensities. Electron/ion temperatures, radiation landscape, and electronic state are sampled at \(9\) different nodes. The transmission and fluorescence spectra were taken from the back node along the forward direction over a \(2\thinspace\pi\) solid angle.}
\end{figure}

We carried out NLTE simulations with a collisional-radiative model to study a \(10\)~\(\mu\)m-thick copper sample that is illuminated by X-rays. We chose a range of fluences (\(5{\times}10^{3}\mbox{--}7{\times}10^{7}\)~J/cm\(^{2}\)) that are relevant in experimental settings of present-day XFELs. The incident beam's time profile \(I_{0}(t)\) was defined as a Gaussian function with \(7\)~fs full width at half maximum (FWHM), centered at \(30\)~fs, with a \(9\)~keV photon energy, and \(\Delta{E}/E=1{\times}10^{-3}\) bandwidth~\cite{madsen_scientific_2013}. Using a screened hydrogenic model, the material was described by a set of energy levels and transition rates for radiative, collisional, and autoionization/electron capture events. Based on the setup shown in figure~\ref{fig:diagram}, we computed the transmission time profile \(I_\mathrm{T}(t)\), fluorescence time profile \(I_\mathrm{F}(t)\), absorption, and occupations resulting from the photo-induced electronic fluctuations. In an experiment, we expect a delay in the radiation path along the thickness of the material that follows the speed of light (approx. \(30\)~fs for 10~\(\mu\)m). In the simulations, radiation is applied instantaneously at each simulation time step with a magnitude that reflects the material's current optical properties along the radiation path.

\section{Results and Discussion}
\subsection{X-ray transmission and fluorescence}
\begin{figure}
    \includegraphics[width=\linewidth]{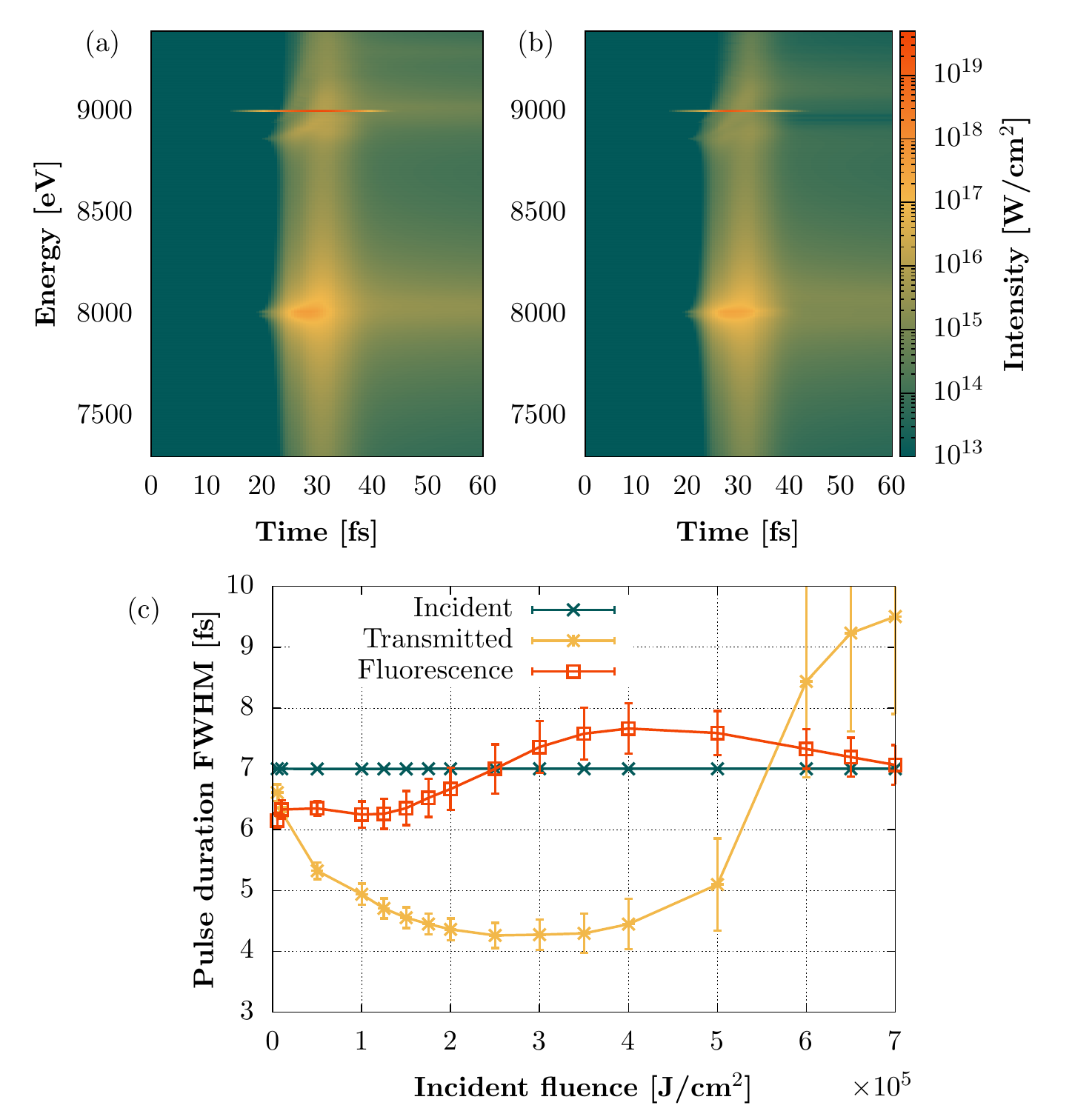}
    \caption{\label{fig:emission}Calculated intensity at the (a)~front and (b)~back of the copper slab irradiated with a \(3.5{\times}10^{5}\)J/cm\(^2\) pulse. We did not consider a specific detector distance and neglected intensity decays following the inverse square law. K\(\alpha_{1}=8012\)~eV, K\(\alpha_{2}=7992\)~eV and K\(\beta_{1}=8868\)~eV. (c) Incident, transmitted (\(9000\)~eV), and fluorescent (\(8006\)~eV) pulse durations with increasing fluence. Error bars represent a \(95\%\) confidence bound of the best fit's width.}
\end{figure}

We initially calculated the duration of the angle-averaged intensity profiles \(I_\mathrm{T}(t)\) and \(I_\mathrm{F}(t)\) as a function of incident fluence. The resulting pulse durations are shown in figure~\ref{fig:emission}(c). The intensity at any node (sampling planes) consists of radiation from two origins: transmission of the X-ray beam and emission from the material. These two contributions along the forward direction made up the detected intensity spectra that, for a single fluence, are shown in figure~\ref{fig:emission}. Panels (a) and (b) correspond to the front and back nodes of the sample, respectively. We defined the fluorescence as the signal that yielded the shortest FWHM and highest peak intensity over the photon energy range between \(7\mbox{--}9\)~keV. We divided the spectra in bins of \(9\)~eV (identical to the \(I_{0}(t)\) bandwidth), computed the aspect ratio as peak intensity/duration, and found \(8006\)~eV to be the strongest. See the \supp for results. We employed a single Gaussian best fit to determine the FWHM.

\begin{figure}
    \includegraphics[width=\linewidth]{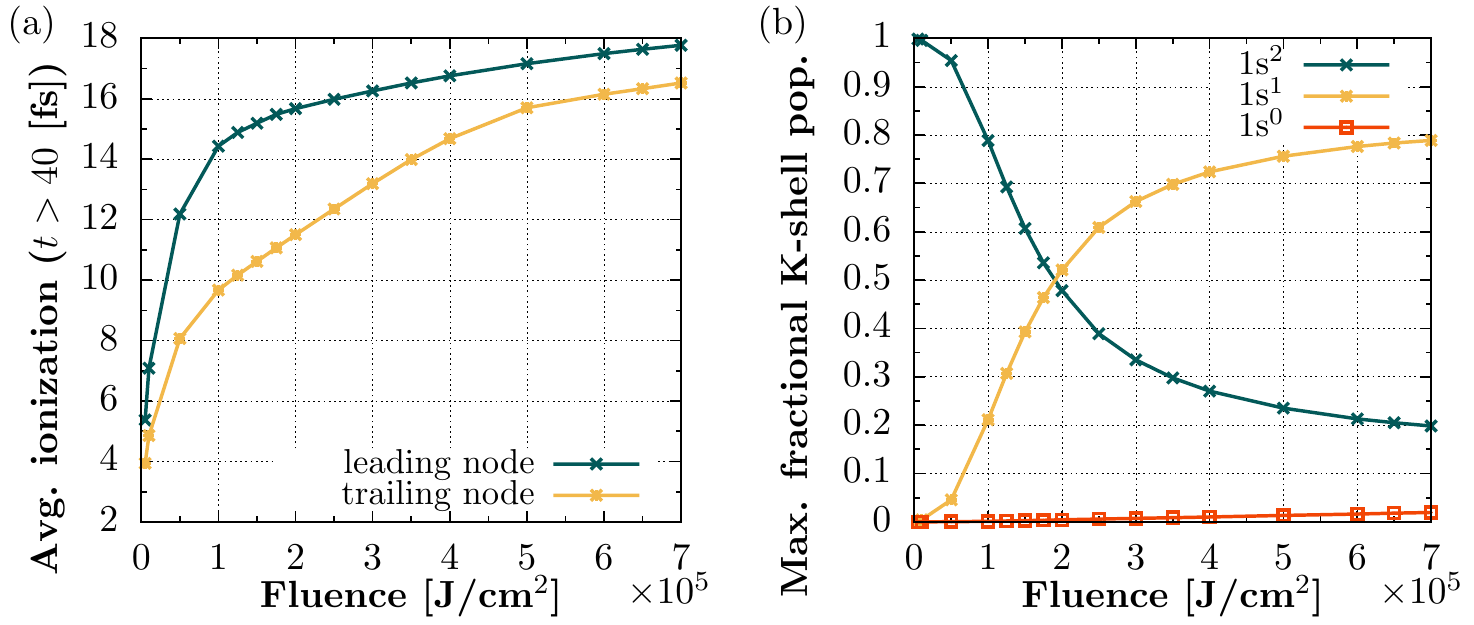}
    \caption{\label{fig:ionzation}(a)~Average ion charge at the end of the pulse and (b)~maximum K-shell population created in the material. The incident photon energy is not large enough to create a second core hole state. We attribute the non-zero \(1s^{0}\) state population to electron-ion collisional ionization. If the rate of this process is faster than the relaxation time of photoionization, a single core hole state can become doubly ionized.}
\end{figure}

The simulated transmission profile FWHM followed experimental results from~\cite{inoue_shortening_2021}, where we found pulse durations of roughly \(4\mbox{--}5\)~fs at fluences of \(2\mbox{--}3{\times}10^{5}\)~J/cm\(^{2}\). We observed some discrepancies at low fluences, where experiments showed pulse times longer or similar to the incident beam. Our simulations instead returned shorter pulse times compared to the incident beam. At higher fluences between \(6\mbox{--}7{\times}10^{5}\)~J/cm\(^{2}\), the simulations predicted longer durations than the incident X-rays. The transmission in these cases featured a double peak that was not well captured by a single Gaussian best fit, resulting in large uncertainty in the FWHM. In the low fluence limit, the final average charge in the copper atoms was below~\(+8\), and the generation of core holes was less than \(10\%\), as shown in figure~\ref{fig:ionzation}. The screened hydrogenic model reliably describes a system with significant ionization but loses accuracy for closed-shell and neutral atoms~\cite{scott_advances_2010}. These artifacts can be corrected by scaling energies to match more detailed calculations~\cite{scott_advances_2010}, but we expect a less accurate system representation in the low ionization regime. The simulations also revealed marginally shorter fluorescence profile FWHM at fluences below \(2.0\)~J/cm\(^{2}\).

\subsection{Temporal suppression mechanism}
\begin{figure}
    \includegraphics[width=\linewidth]{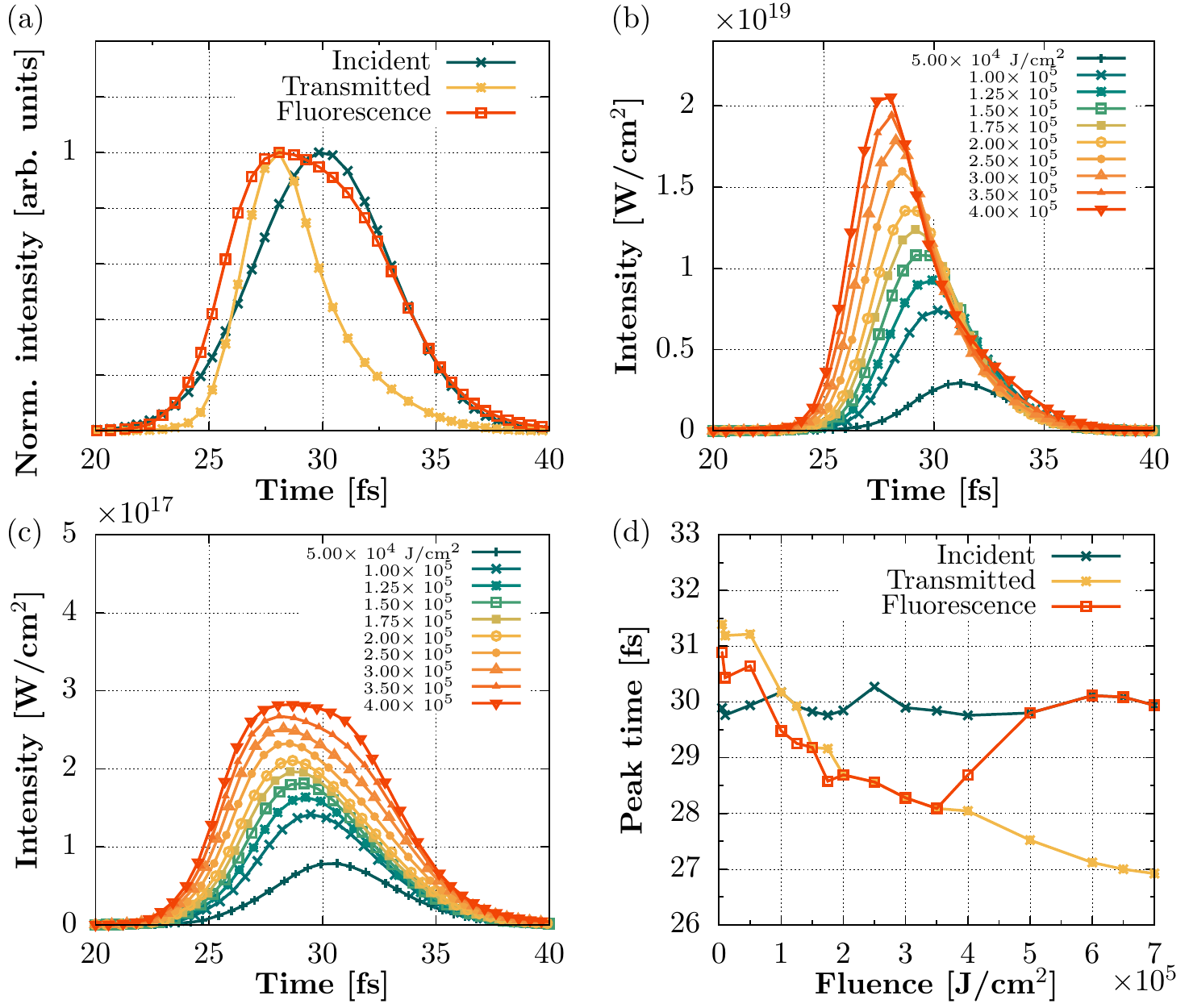}
    \caption{\label{fig:intensity}(a)~Radiation dynamics for a \(3.5{\times}10^{5}\)~J/cm\(^{2}\) incident beam revealed transmission occurred early in the radiation exposure and extinguished before the peak of the incident pulse at \(30\)~fs. (b)~Transmission profiles shifted earlier in time with increasing fluence and (c) emission at \(8006\)~eV became wider with increasing fluence. (d)~Summary of peak intensity.}
\end{figure}

To understand the calculated \(I_\mathrm{T}(t)\) and \(I_\mathrm{F}(t)\) intensity profile durations with a \(9\)~keV incident beam, we explored the dynamics of the transmission and fluorescence relative to the initial pulse. Figure~\ref{fig:intensity}(a) shows normalized profiles for a single fluence of \(3.5{\times}10^{5}\)~J/cm\(^{2}\). We found transmission peaked and died out earlier than the incident signal, while the fluorescence persisted over the entire duration of the incident signal. Figures~\ref{fig:intensity}(b) and (c) generalize these results, displaying \(I_\mathrm{T}(t)\) and \(I_\mathrm{F}(t)\) for varying fluences. For transmission, transparency and termination tended to happen at earlier times as fluence increased. Fluorescence FWHM increased with increasing fluence and peak times shifted earlier in time at fluences below \(3.5{\times}10^{5}\)~J/cm\(^{2}\) and shifted to later times at higher fluences. Peak times for all three profiles are summarized in figure~\ref{fig:intensity}(d).

\begin{figure*}
    \centering
    \includegraphics[width=0.75\linewidth]{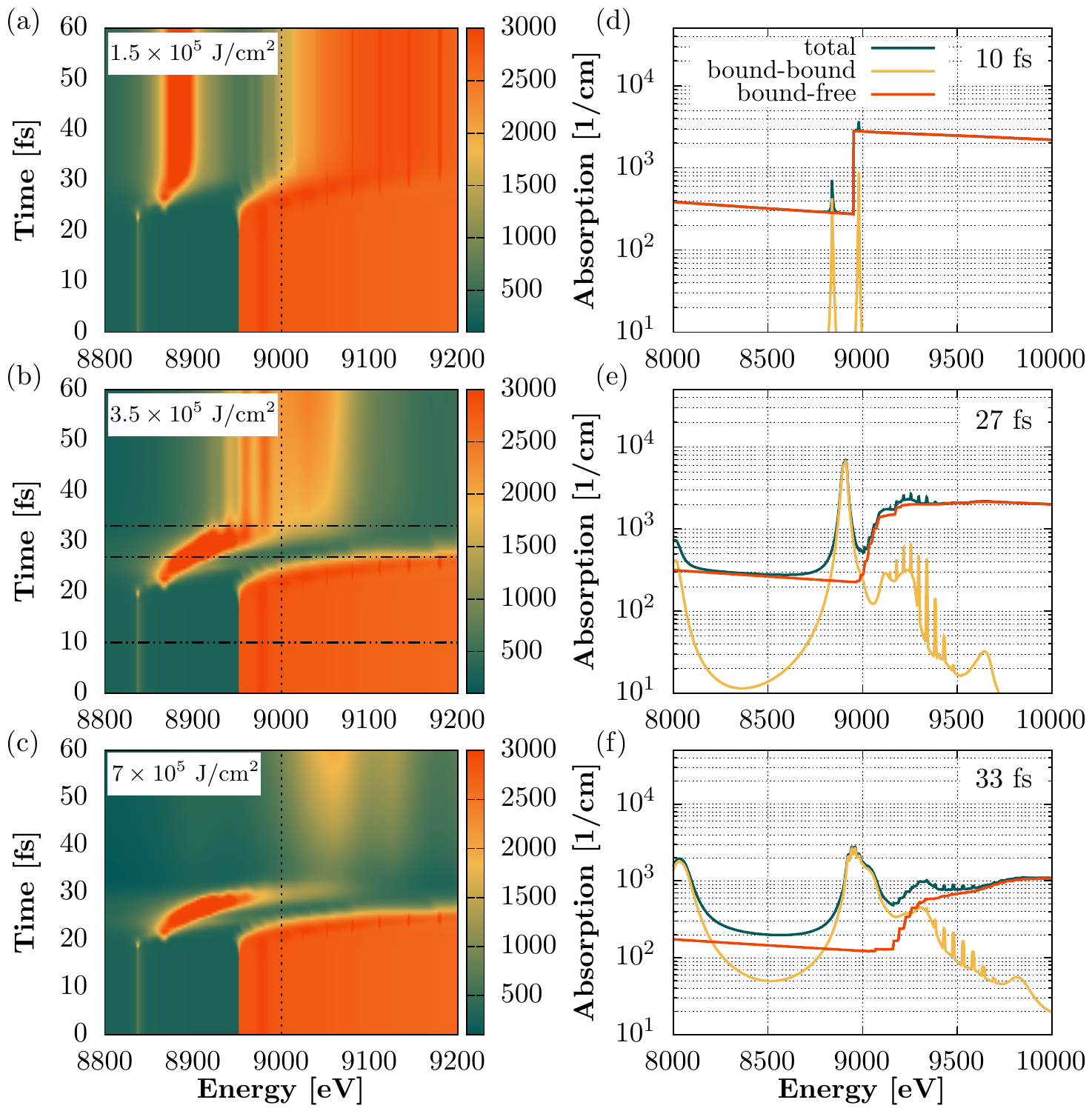}
    \caption{\label{fig:absorption} (a)-(c)~Opacity averaged over all zones showing displacement of the K-edge and resonant \(K\mbox{--}M\) transition for three incident fluences. The vertical dashed line indicates the XFEL pulse photon energy. Horizontal lines in (b) are cuts through the absorption shown in the three panels to the right. (d)-(f)~Opacities at \(3.5{\times}10^{5}\)~J/cm\(^{2}\) averaged over all zones at three instances during the simulation. The \(10\)~fs snapshot shows opacities for a cold sample, the \(27\)~fs shows a dip in the opacity at \(9\)~keV, and the \(33\)~fs snapshot shows an increased opacity at \(9\)~keV.}
\end{figure*}

\paragraph{Transmission profile}
We found saturable absorption offered an incomplete description of transmission profiles. When absorption saturates, the transmitted X-rays should match the incident pulse. Instead, our calculations revealed transmission terminated well before the incident beam. Figures~\ref{fig:absorption}(a)-(c) display the absorption coefficient of the material near the copper K-edge as a function of time for fluences of \(1.5\), \(3.5\), and \(7{\times}10^{5}\)~J/cm\(^{2}\), respectively. In all cases, we observed shifts in the edge plus an opaque feature at photon energies below the edge corresponding to a \(K\mbox{--}M\) transition. We found the most dominant contribution at \(9\)~keV came from a \(1s\mbox{--}3p\) transition, where the Cu atoms reached ionization levels between \(+9\) to \(+17\). For low fluences, the shortening of the transmission profile duration was uniquely a consequence of frustrated absorption. The initial section of the beam was absorbed until the K-edge moved to larger energies. For sufficiently high fluences, the opaque transition shifted into the photon energy range of the incoming X-rays, effectively forming a transient transparent window in the material. The outcome was an even shorter transmission. At more extreme fluences, the resonant transition shifted into the photon energy range of the incoming X-rays but was promptly suppressed by the sheer number of incident photons resulting in a double peak profile with a large FWHM.

We believe the reason for the resonant \(K\mbox{--}M\) state's proliferation and motion along the path of the beam is similar to that of photon energy shifts in emission spectra for high-temperature plasma discussed in literature~\cite{chung_generalized_2016,vinko_creation_2012,young_femtosecond_2010}. The main mechanism for absorption is K-shell ionization resulting in a single core hole. K-shell fluorescence and electron ejection are competing processes with comparable probabilities. In copper, fluorescence accounts for \(44.5\%\) of the total recombination while the remaining holes are filled mainly via \(KLL\) Auger-Meitner decay~\cite{bambynek_X-ray_1972}. Electron impact is another source of ionization. Hot electrons ejected by collision with the X-ray beam or through Auger-Meitner decay generate further vacancies in the material, triggering an ionization cascade. Primary and secondary ejected electrons equilibrate through collisions with the cold electron reservoir (conduction band). Cold electrons also gain kinetic energy and begin to ionize outer valance states in the material. As more bound electrons exit the atoms, screening of higher levels is reduced, and deep states move closer to the nucleus. For high enough charged states the \(1s\mbox{--}3p\) transition (most dominant around \(9\)~keV) increases and shifts into the range of the incident beam. Photo-electrons are no longer ejected to the continuum and are instead resonantly pumped to the \(M\)-shell~\cite{young_femtosecond_2010}. The final transmitted profile duration is determined by the time delay between the K-edge and \(1s\mbox{--}3p\) photon energy changes along the thickness of the sample.

\begin{figure}
    \includegraphics[width=\linewidth]{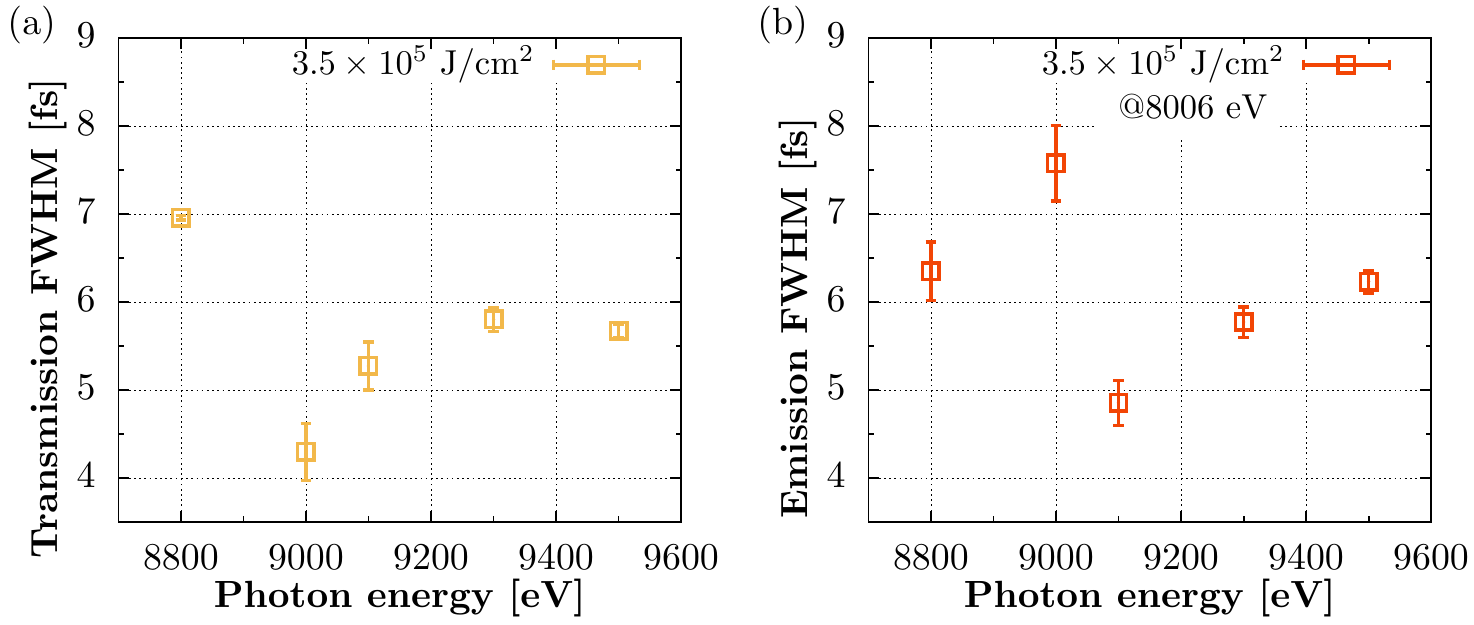}
    \caption{\label{fig:pulse_duration_ev}(a)~Transmission and (b)~fluorescence with varying beam photon energies. At \(9\)~keV, fluorescence is extended by the resonant \(K\mbox{--}M\) transition. At \(9.1\)~keV the fluorescence signal is temporally shorter because the resonance does not extend to this photon energy. See \supp for full spectra comparison.}
\end{figure}

\paragraph{Fluorescence profile}
For a given fluence we fitted a linear combination of the normalized fluorescence profile given by the simulations using \(\hat{I}_\mathrm{F}(t)=a_{1}\thinspace\hat{I}_\mathrm{0}(t)+a_{2}\thinspace\hat{I}_\mathrm{T}(t)\), where \(a_{1}\) and \(a_{2}\) are coefficients that changed based on the incident fluence. We found \(\hat{I}_\mathrm{F}(t)\) at low fluence was mainly described by the transmission profile and at high fluence by the incident profile. More information is found in the \supp. These changes in the coefficients suggested absorption caused by the \(K\mbox{--}M\) resonance extended the temporal duration of the resulting K\(\alpha\) emission. We expect fluorescence to follow the incident profile for a linear material response. In a non-linear regime, the emission's duration changes at different fluences owing to variations in the opacity. The effects of nonlinearity became apparent at \(1.5{\times}10^{5}\)~J/cm\(^{2}\). Figure~\ref{fig:pulse_duration_ev} shows the result of modifying the photon energy of the incoming X-rays. The lowest fluorescence profile FWHM occurred at an incident photon energy of \(9.1\)~keV, where the incident beam completely avoided the resonance. These results indicate the beam's photon energy could be adjusted to minimize fluorescence duration.

\subsection{Limitations of the model}
We are interested in changes to the radiation and copper's electronic population along the beam's trajectory. The collisional-radiative code distributes radiation instantaneously but follows the optical properties of the material. A \(9\)~keV X-ray beam incident at the front of a cold copper sample instantly appears along the entire thickness, but its magnitude is significantly reduced due to absorption. Emission spreads isotropically but is 1 or 2 orders of magnitude smaller than the incident beam and has a diminished impact on copper's state. Compared to a model that follows the speed of light, we expect the approximate treatment of the radiation to minimally influence the radiation landscape and state of the material. A graphical description of instant propagation is found in the \supp.

The collisional-radiative code also assumes photo-ionized electrons instantly thermalize. The energy distribution, which would otherwise be comprised of a thermal and non-thermal contribution~\cite{hau-riege_nonequilibrium_2013}, remains Maxwellian. As a consequence, the model predicts a greater number of thermal electrons at a higher temperature than a model that considers a bimodal distribution~\cite{hau-riege_nonequilibrium_2013}. Primary ejected electrons cause a cascade of electron impact events that determine electronic, optical, and radiative material changes. For copper, the absolute collision ionization cross-section grows with increased electron temperature, peaking at approximately \(40-50\)~eV and falls slowly at larger temperatures~\cite{lotz_electron-impact_1969}. Highly energetic electrons (\(<80\)~eV) are likely to ionize from deep valance shells, while lower-energy electrons ionize from outer valance shells~\cite{bartlett_calculation_2002}. The average electron temperature in a typical simulation performed in this study can reach several hundreds of eV. For the first few femtoseconds, our simulations could underestimate the secondary ionization of deep valance states while overestimating outer valance electron ionization. When the electron temperature reaches several hundred eV the simulations could underestimate the outer valence electron ionization. The shift in the resonant state is sensitive to the ionization level of the system~\cite{vinko_creation_2012}. We expect the secondary ionizations to balance out during the simulations. To fully evaluate the effects of electron impact ionization it is necessary to compare results with collisional-radiative codes that evolve the non-thermal electron distribution~\cite{fontes_alamos_2015,macfarlane_simulation_2004,le_conservative_2017}.

\section{Conclusions and Outlook}
We calculate exposure to high-ionizing radiation on a thin copper sample can be used to temporally decrease the temporal duration of an X-ray pulse via transmission and fluorescence. Starting with a \(7\)~fs FWHM beam at \(9\)~keV, we found approximately a \(3\)~fs reduction in transmission. Opacity calculations revealed saturable absorption was partially responsible for the temporal reduction. At sufficiently high intensities shifts in the \(K\mbox{--}M\) resonant excitation turned the copper plasma opaque, causing an early transmission termination. The \(K\mbox{--}M\) shift also caused extended fluorescence with longer FWHM than the incident beam. By increasing the photon energy we escaped this resonance and achieved as much as a \(2\)~fs reduction via fluorescence at \(8006\)~eV.

We propose the present work can be expanded in three ways:
\vspace{-\topsep}
\begin{itemize}
    \setlength{\parskip}{0pt}
    \setlength{\itemsep}{1pt}
    \item[(1)] We hypothesize a two-colour scheme, above and below the Cu K-edge with femtosecond time delay, can be used to control the transmission profile FWHM and enhance photon yield. The first signal above the edge could trigger a shift in the resonant transition, while the second beam could propagate largely unattenuated until the \(K\mbox{--}M\) resonance crosses its path.
    \item[(2)] An approach to reach shorter profile durations is to use a copper alloy to decrease the time it takes for the resonant \(K\mbox{--}M\) transition to shift to higher energies. For example, nickel has a higher Auger-Meitner yield and can strongly interact with Cu K\(\alpha\). A Cu-Ni mixture augments the number of electrons available for electron collision ionization, potentially reaching a high degree of ionization in copper faster.
    \item[(3)] We can study the effects of modifying the X-ray interaction with matter on saturable absorption by either increasing the beam's photon energy to induce a doubly ionized core (decay channels might be different) or by utilizing nano-structured targets~\cite{hollinger_efficient_2017}.
\end{itemize}

The short pulses predicted in this work or with the above-proposed approaches could be useful for incoherent diffraction imaging of metalloproteins, where the speckle pattern visibility inversely depends on the number of coherent intervals of the fluorescing atoms~\cite{goodman_statistical_2015}. Signal reaching the detector from a single exposure with FWHM will contain approximately FWHM\(/\tau_\mathrm{c}\) number of modes where \(\tau_\mathrm{c}\) is the fluorescence coherence time~\cite{trost_photon_2020}. A shorter pulse can help preserve contrast in the speckle pattern by reducing the number of modes.

\section{Methods}
We performed one-dimensional non-local thermodynamic equilibrium (NLTE) simulations with collisional-radiative code \cretin~v\_02\_20~\cite{scott_cretinradiative_2001} and a screened hydrogenic model (SHM) based on principal quantum number. We included solid-density effects via electron degeneracy and continuum lowering to describe a solid-to-liquid transition~\cite{deschaud_atomic_2015}. We recorded time-varying radiative properties (opacities and emissivities), material properties (temperature, densities, population states), and detailed radiation spectra. The physical choices presented in this section were made through commands inside the code and their implementation can be found in~\cite{scott_glf_1994,scott_cretinradiative_2001}. \cretin has been compared both with other NLTE codes~\cite{hansen_review_2020} and with experiments that study warm dense matter originating from the interaction of proteins, water and metals with soft and hard XFEL beams~\cite{beyerlein_ultrafast_2018, barty_self-terminating_2012, andreasson_saturated_2011}.

As a starting configuration, we defined copper as a degenerate and strongly coupled plasma with a fixed density of \(8.96\)~g/cm\(^{3}\) and temperature of \(0.025\)~eV (\(290\)~K), see \supp for details. At standard temperature and pressure, solid copper is treated on average as a pseudo-noble gas electronic configuration \([Ar]3d^{10}\) with its \(4s\) electron occupying the conduction band. We modelled this band by placing one electron per atom in the continuum. We fixed the plasma starting thermal conductivity to match the copper conductivity at \(20\)~\(^{o}\)C and \(1\)~bar of \(3.86{\times}10^{7}\)~ergs/cm\(^{2}\)/s~\cite{carvill_3_1993}. Simulations ran for \(60\)~fs in dynamic steps of \({\approx}0.5\)~fs. Cold opacities are directly calculated from the atomic model.

\subsection{Collisional-radiative algorithm}
The collisional-radiative algorithm solves atomic kinetics, radiation transport, density, and temperature equations. For NLTE conditions, the radiation depends on knowledge of the populations, which in turn are prescribed by the radiation, and the solution is reached self-consistently~\cite{carlsson_radiative_1998}. The first kinetics iteration is based on an initial guess of the populations to establish the spectroscopic properties of the material. Radiation transfer supplies heating rates for temperatures, which are used to calculate new densities. Densities influence the kinetics, and the loop repeats iteratively until populations converge. In the following time step, the incident radiation is modified and the loop is reiterated.

\paragraph{Atomic kinetics.}
\cretin solves atomic kinetics using the rate equation \(d\mathbf{y}/dt=\mathbf{A}\thinspace\mathbf{y}\), where \(\mathbf{A}\) is the rate matrix and \(\mathbf{y}\) is the population of the atomic levels~\cite{scott_glf_1994}. The rate matrix contains transitions for processes included in the atomic model, which are adjusted based on the density, electron distribution, and incident radiation field. The sample is divided into nodes where atomic kinetics are determined independently based on the local environment~\cite{scott_glf_1994}. Material cannot move across boundaries and energy exchange between nodes is accomplished via radiation transport and thermal conduction~\cite{scott_glf_1994}. The populations dictate opacities and emissivities, which are passed as inputs to the radiation transfer algorithm.

\paragraph{Radiation transfer.}
\cretin keeps track of the changing energy landscape with a radiative transfer equation that, for a frequency \(\nu\)-dependent field \(I_{\nu}\) travelling over a straight path \(s\) through the sample, is written as \(d I_\mathrm{\nu} / d\tau_\mathrm{\nu} = I_\mathrm{\nu}-S_\mathrm{\nu}\)~\cite{carlsson_radiative_1998}. Here \(S_{\nu}\equiv j_{\nu}/\alpha_{\nu}\) is the source function defined as the ratio between the emissivity \(j_{\nu}\) and extinction coefficient \(\alpha_{\nu}\) of the sample, and \(\tau_{\nu}\) is the optical depth along the infinitesimal path \(ds\) defined as \(d\tau\equiv-\alpha{\thinspace}ds\)~\cite{carlsson_radiative_1998}. The radiative transfer treatment includes paths in multiple directions plus the symmetries inherent in the one-dimensional geometry to cover the total solid angle.

For numerical efficiency, radiation is handled using independent energy spaces with unique grid sizes and ranges. A continuum space is used to evaluate photoionization and photoexcitation integrals that couple to the atomic kinetics, and a spectral space is used to construct high-resolution spectra based on real-time plasma conditions~\cite{scott_cretinradiative_2001}. To reduce computational demands on the continuum integrals, we defined a coarse mesh over the photon energy range \(0.1\)~eV--\(10\)~keV. We identified optically-thick transitions and checked the continuum adequately matched the opacity spectrum (see the \supp for details). A formal solution to the radiation transfer problem for continuum was obtained using the Feautrier formalism~\cite{feautrier_sur_1964}. We considered Stark broadening effects when generating the spectra. To model self-absorption, we used escape factors that interpolate between tabulated values~\cite{apruzese_analytic_1985} valid for a static material and the Sobolev limit applicable to fast-expanding material. We also included Compton scattering with a \(1+\cos(\theta)^{2}\) dipole angular dependence.

\paragraph{Temperature evolution.}
Free electrons and ions follow a Maxwell distribution with temperatures evolved from the coupled differential equations
\begin{align}
    \frac{\mathrm{d} T_\mathrm{e}}{\mathrm{d}t}=&\thinspace \frac{2}{3\thinspace n_\mathrm{e}}\left(R_\mathrm{a}+\frac{\mathrm{d}\kappa_\mathrm{e}}{\mathrm{d}x}\frac{\mathrm{d}T_\mathrm{e}}{\mathrm{d}x}\right) - \frac{T_\mathrm{e}}{n_\mathrm{e}}\frac{\mathrm{d}n_\mathrm{e}}{\mathrm{d}t} + \\ \nonumber & \thinspace + \gamma_\mathrm{ei}\left(T_\mathrm{i} - T_\mathrm{e}\right)+S_\mathrm{e}\ ,\\
    \frac{\mathrm{d}T_\mathrm{i}}{\mathrm{d}t}=&\thinspace\gamma_\mathrm{ie}\left(T_\mathrm{e}-T_\mathrm{i}\right)+S_\mathrm{i}\ ,
\end{align}
where \(T_\mathrm{e}\) and \(T_\mathrm{i}\) are the electron and ion temperatures, \(n_\mathrm{e}\) is the local electron density, \(R_\mathrm{a}\) is the heating rate from atomic kinetics, \(\kappa_\mathrm{e}\) the electron thermal conductivity, \(\gamma_\mathrm{ei}\) the electron-ion coupling, and \(S_\mathrm{e}\) and \(S_\mathrm{i}\) are electron and ion source functions for laser absorption~\cite{scott_glf_1994}. The atomic heating rate includes electron/photon-induced ionization or recombination, electron collisional excitations or deexcitation, autoionization, electron capture and Bremsstrahlung~\cite{scott_glf_1994}. We used electron thermal conduction coefficients by~\citet{lee_electron_1984} with a solid-density asymptote. Collisions in plasma are mediated by short and long-distance interactions (hard and soft collisions). The relative cross-sections between these two collisional modes are specified by the Coulomb logarithm from~\citet{brown_temperature_2007}, which also controls Bremsstrahlung and the source function for laser absorption. As a consequence of this temperature formulation, photo-ejected electrons (folded in the heating rate) instantly thermalize with the continuum.

\subsection{Screened hydrogenic data}
A good description of energy levels and transition rates dictates the accuracy of material properties, radiation transport, and spectroscopic features. A problem-specific model based on self-consistent quantum calculations requires significant computational effort to construct if the material is highly ionized, while a general SHM requires less time to define and maintains good accuracy for the intended application if inclusive of all configurations involved in the atomic kinetics to produce accurate spectra~\cite{rubiano_screened_2002,hansen_balancing_2016}. However, using any single set of screening coefficients produces systematic inaccuracies in level energies which become worse around closed shell ions. A SHM for copper that includes quantum numbers for all possible ionization states can yield a large number of transitions that quickly becomes incomputable. We limited ourselves to generate data based on principal quantum number (PQN) \(N\) following methods described by~\citet{scott_advances_2010} and from a convergence study summarized in the \supp.

We defined \(N=24\) energy levels for each atomic state using screening constants from~\citet{more_electronic_1982}, scaled to match ionization energies from quantum calculations by~\citet{liberman_self-consistent-field_1994}, and allowed a maximum of \(5\) possible excitations to to the highest \(N\). We also split photo-induced bound-bound transitions between PQNs for up to \(N=6\) for each atomic state and applied an additional width over each transition to represent fine structure details. \citet{scott_advances_2010} showed the above approach improves the accuracy and distribution of transition energies for xenon resulting in similar spectra compared to that obtained from more sophisticated models.

Rates from photon, electron and ion collisions are also part of the atomic model. Relevant processes included are (1)~photon or electron-induced ionization and recombination, (2)~photon or electron-induced excitation and deexcitation, and (3)~autoionization and electron capture. \cretin computes photo-induced transitions and ionizations based on oscillator strengths from screening constants. We used collisional excitations rates from \jjatom~\cite{chung_fast_2007}, collisional ionizations from~\citet{golden_ionization_1977}, and autoionization from~\citet{chung_flychk_2005}. Rates are influenced by the radiation field and free-electron density. Finally, we include ionizations via the collision of slow-moving highly-charge ions with neutral atoms using a classical overbarrier charge exchange model~\cite{sattin_classical_2000}.

\subsection{Solid density effects}
The atomic data used in Cretin is most appropriate for low density plasmas and is most readily applied with Maxwellian electron distributions to calculate transition rates and material properties. At low temperatures and solid density, electrons instead follow a Fermi-Dirac distribution \(F_\mathrm{e}\). Degeneracy was implemented in the code in the following manner. Density and pressure were modified to model a degenerate electron gas and excitation rates were multiplied with a factor that adds degeneracy effects~\cite{scott_collisional-radiative_2016}. Transitions that involve free-electrons, such as collisional ionization, required Pauli-blocking factors \(P(\epsilon)=1-F_\mathrm{e}(\epsilon)\) based on electronic occupation~\cite{scott_collisional-radiative_2016}.

The atomic data is defined for an isolated atoms, where we expect the number of energy levels based on PQN for each charge state to grow large near the continuum~\cite{scott_collisional-radiative_2016}. If we now consider the environment, the energy required to ionize a bound electron is lowered by the electrostatic potential of neighbouring atoms and free electrons. The existence of Coulomb interactions also modifies the free energy which generally contributes to a negative pressure~\cite{scott_collisional-radiative_2016}. Continuum lowering cuts the number of available PQN states and shifts rates, thus altering the opacity and thermodynamics of the system~\cite{ciricosta_direct_2012}. We employed the Ecker and Kröll~\cite{ecker_lowering_1963} continuum lowering model and motivate our choice based on experimental findings that measured the ionization state of solid-density aluminum from K-alpha fluorescence emission~\cite{ciricosta_direct_2012,preston_effects_2013}. Ecker-Kröll was shown to correctly estimate continuum lowering in high-charged states and predict K-edge shifts under conditions similar to those in these experiments~\cite{vinko_density_2014,crowley_continuum_2014}.

The effect of continuum lowering can change dynamically during the simulation. The code gradually reduces each atomic level's statistical weight \(W\) using a smooth function. To calculate the lowering weight on each charge state, the code employs the expression
\begin{equation}\label{eq:degeneracy_lowering}
    W = \exp{\left[\frac{\Delta E_{\text{max}} - \Delta E}{\Delta E_{\text{max}}}\right]^{\gamma}}\ .
\end{equation}
Here \(\Delta E_{\text{max}}\) represents the energy that is needed to make the state disappear and \(\Delta E\) is the degeneracy lowering calculated for the current plasma conditions. The predicted ionization states at various temperatures depends on degeneracy and continuum lowering~\cite{ali_improved_2018}. The parameter \(\gamma\) was set to 2 and the calculation of \({\Delta}E\) used an ion sphere model that matched copper's conduction band at low temperature and solid density.

\begin{acknowledgments}
    The authors would like to thank Ibrahim Dawod, Christina Vantaraki, Dr.\ Hai P.\ Le for discussions, and Filipe Maia for providing the computing resources for simulations. S.\ C.\ and N.\ T.\ thank the Swedish Research Council (Grant 2019-03935) for financial support. C.\ C.\ acknowledges the Swedish Research Council (Grant 2018-00740) and the Helmholtz Association through the Center for Free-Electron Laser Science at DESY. The work of H.A. Scott was performed under the auspices of the U.S. Department of Energy by Lawrence Livermore National Laboratory under Contract DE-AC52-07NA27344.
\end{acknowledgments}
\bibliographystyle{apsrev4-2}
\bibliography{ms}
\end{document}


\title{Decreasing ultrafast X-ray pulse durations with saturable absorption and resonant transitions}
\author{Sebastian Cardoch}
\email{sebastian.cardoch@physics.uu.se}
\affiliation{Department of Physics and Astronomy, Uppsala University, Box 516, SE-751 20, Uppsala, Sweden}
\author{Fabian Trost}
\affiliation{Center for Free-Electron Laser Science, Deutsches-Elektronen Synchrotron (DESY), Hamburg, Germany}
\author{Howard A. Scott}
\affiliation{Lawrence Livermore National Laboratory, L-18, P.O. Box 808, 94550, Livermore, CA, USA}
\author{Henry N. Chapman}
\affiliation{Center for Free-Electron Laser Science, Deutsches-Elektronen Synchrotron (DESY), Hamburg, Germany}
\affiliation{The Hamburg Center for Ultrafast Imaging, Universität Hamburg, Luruper Chaussee 149, 22761 Hamburg, Germany}
\affiliation{Department of Physics, Universität Hamburg, Luruper Chaussee 149, 22761, Hamburg, Germany}
\author{Carl Caleman}
\affiliation{Department of Physics and Astronomy, Uppsala University, Box 516, SE-751 20, Uppsala, Sweden}
\affiliation{Center for Free-Electron Laser Science, Deutsches-Elektronen Synchrotron (DESY), Hamburg, Germany}
\author{Nicusor Timneanu}
\email{nicusor.timneanu@physics.uu.se}
\affiliation{Department of Physics and Astronomy, Uppsala University, Box 516, SE-751 20, Uppsala, Sweden}
\date{\today}
\maketitle
\section{\supp}
\newpage
\subsection{Determining optimal fluorescing photon energy range}
\begin{figure}[ht]
    \includegraphics[width=\linewidth]{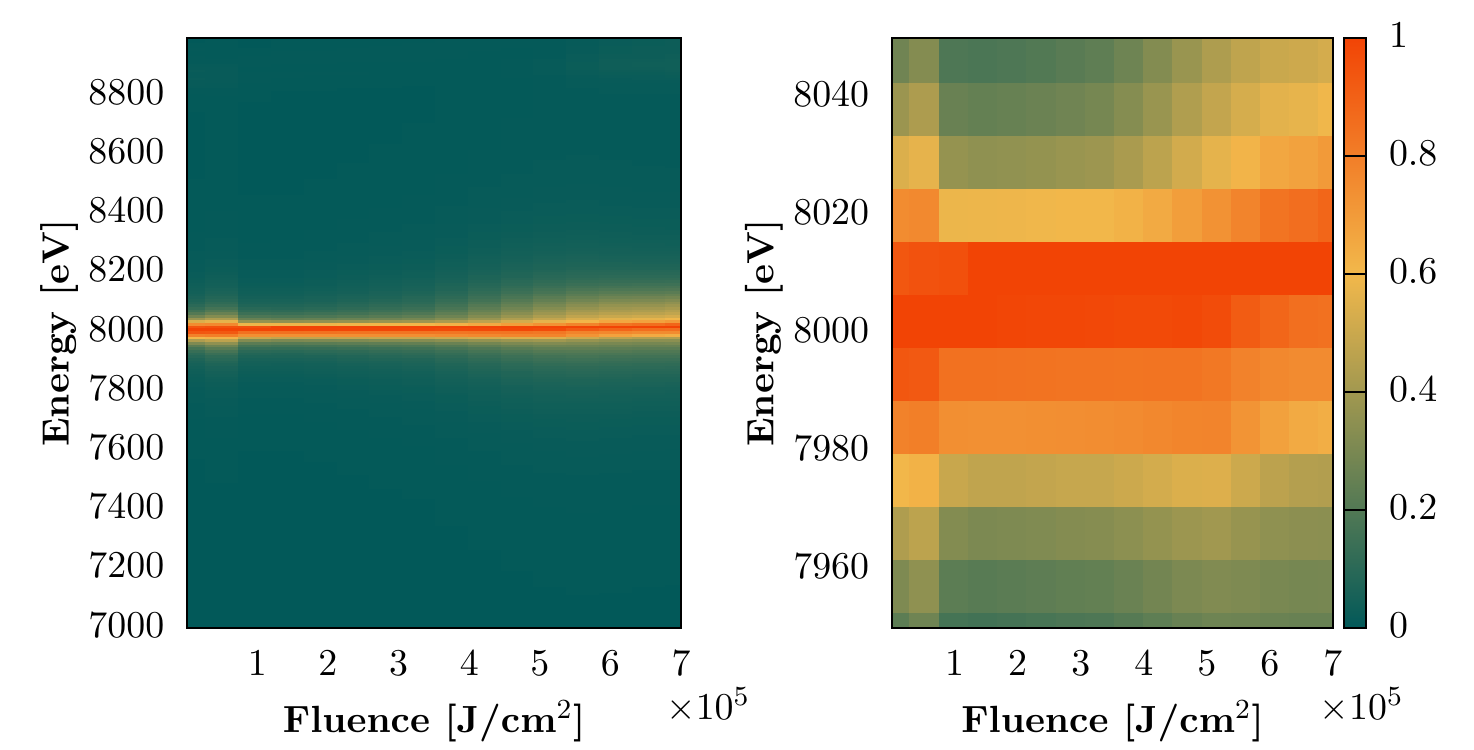}
    \caption{\label{fig:apect_ratio}Aspect ratio of emission pulses at various photon energies over a range of incident fluence values. The aspect ratio has been normalized---at each fluence---to be over the range \(0\) to \(1\). An aspect ratio close to 1 translates to a short pulse with high intensity, while an aspect ratio close to 0 translates to a wider pulse with a lower intensity.}
\end{figure}

To determine the optimal fluorescence signal we swept over the energy spectrum in bin sizes of \(9\)~eV between \(7002\mbox{--}9000\)~eV. At each selected energy interval, we averaged the intensity within the bandwidth, determined the profile's maxima, and fitted a Gaussian function to resolve the FWHM. We computed the line profile's aspect ratio and, to compare over all fluences, normalized the results at each fluence. Based on plots shown in figure~\ref{fig:apect_ratio}, at low fluence we found an emission photon energy between K\(\alpha_{1}\) and K\(\alpha_{2}\) to be optimal. At high fluence, the optimal emission photon energy shifted to higher photon energy values. We selected the interval \(8006\mbox{--}8015\)~eV.

\newpage
\subsection{Modelling the fluorescence profile}
\begin{figure}[ht]
    \includegraphics[width=\linewidth]{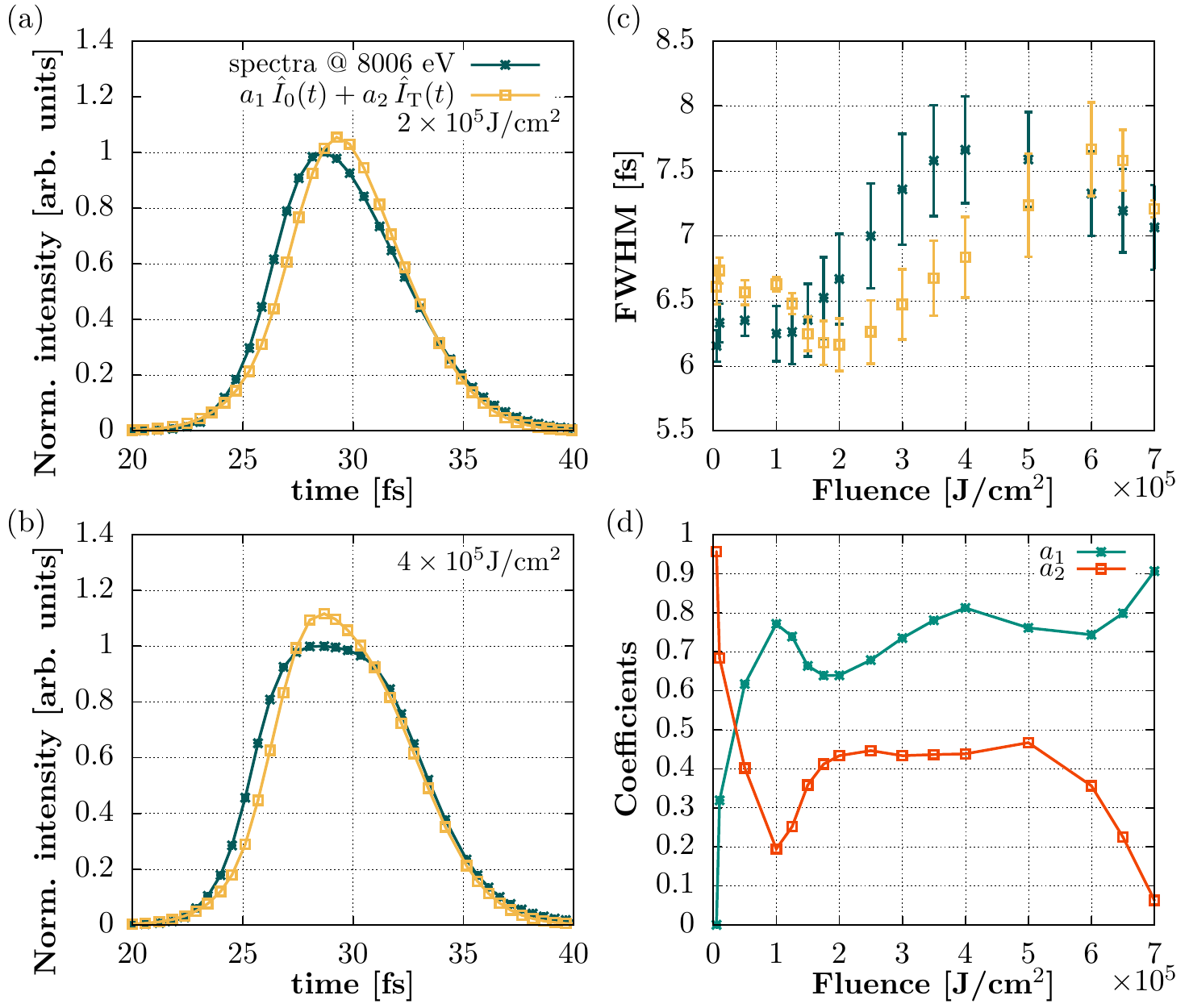}
    \caption{\label{fig:fluorescence_approximation}(a)(b)~Comparing simulated and fitted normalized fluorescence profiles for two different fluences. One profile is taken from simulations and the other is the result of fitting a linear combination. (c)~Comparing FWHM from spectra-derived and linear combination calculated as a function of incident fluence. We computed the linewidth with a fitted Gaussian function. The error represents a \(95\%\) confidence bound of the best fit's width. (d)~The coefficients were worked out by solving the linear system of equations \([I_\mathrm{0}(t)\thickspace I_\mathrm{T}(t)]\thinspace X=I_\mathrm{F}(t)\), where \(X=[a_{1}\thickspace a_{2}]\), using a non-negative least squares curve fitting solver.}
\end{figure}

\newpage
\subsection{Comparing spectrum at two incident photon energies}
\begin{figure}[ht]
    \includegraphics[width=\linewidth]{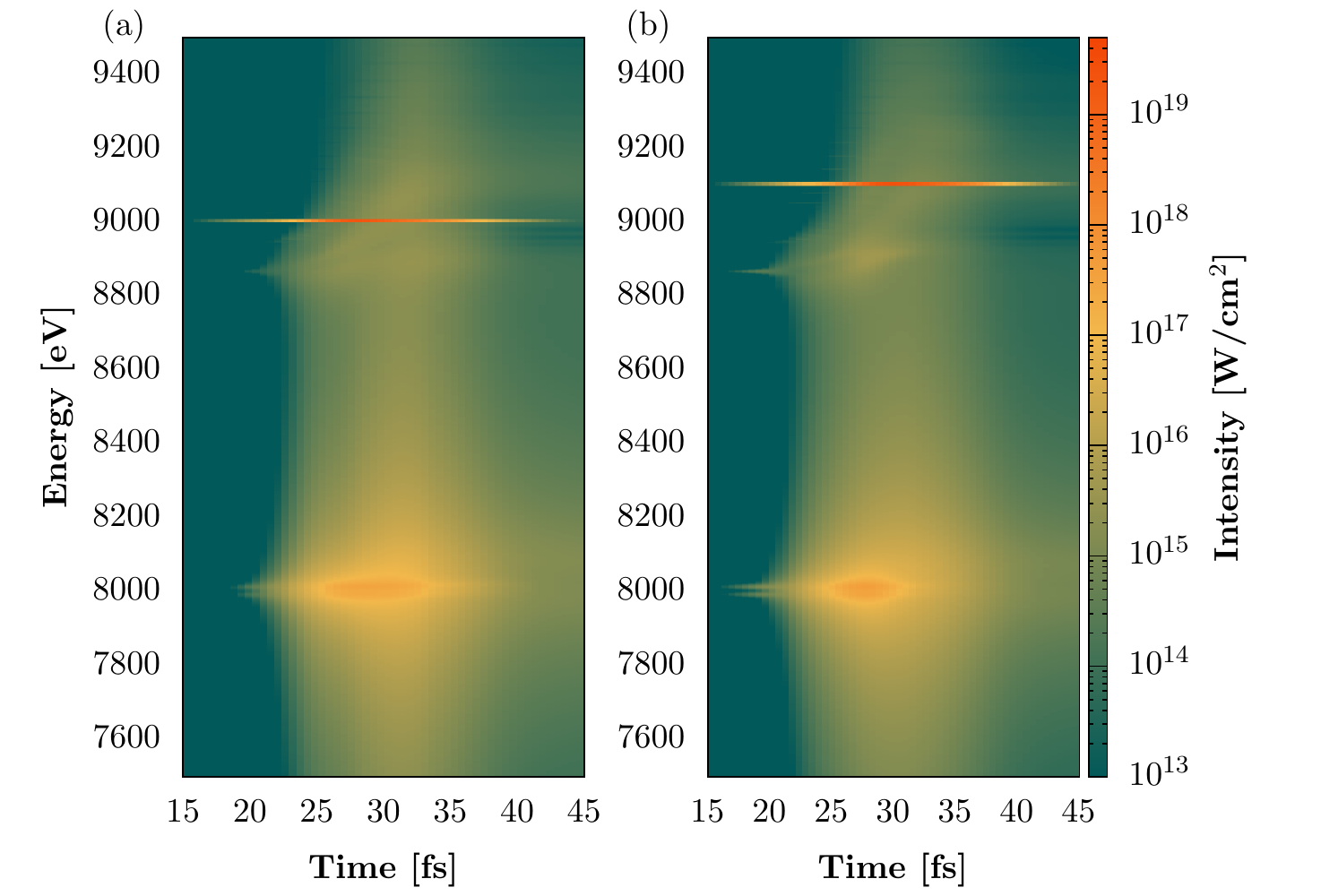}
    \caption{\label{fig:pulse_ev_spectra} Spectra from the back of the sample when irradiated with \(3.5{\times}10^{5}\)~J/cm\(^{2}\) fluence at (a) \(9\)~keV and (b) \(9.1\)~keV.}
\end{figure}

\newpage
\subsection{Instantaneous radiation propagation}
\begin{figure}[ht]
    \includegraphics[width=0.7\linewidth]{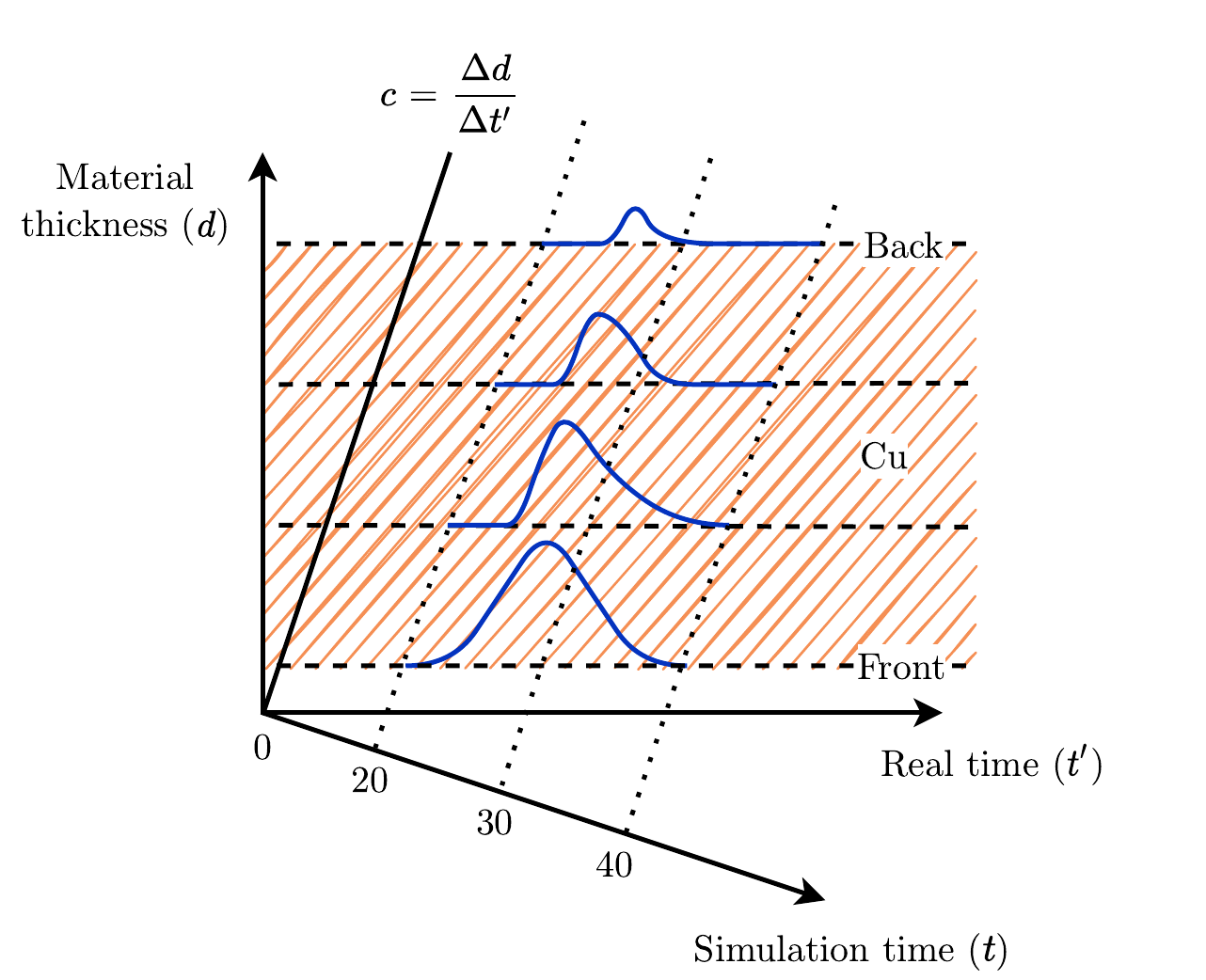}
    \caption{\label{fig:timediagram}Comparing the propagation of the incident radiation along the thickness of the copper between an experiment to the simulation environment. Blue lines indicate the radiation time profiles at \(9\)~keV photon energy at different thicknesses in the material.}
\end{figure}

\newpage
\subsection{Material optical depth}
\begin{figure}[ht]
    \includegraphics[width=\linewidth]{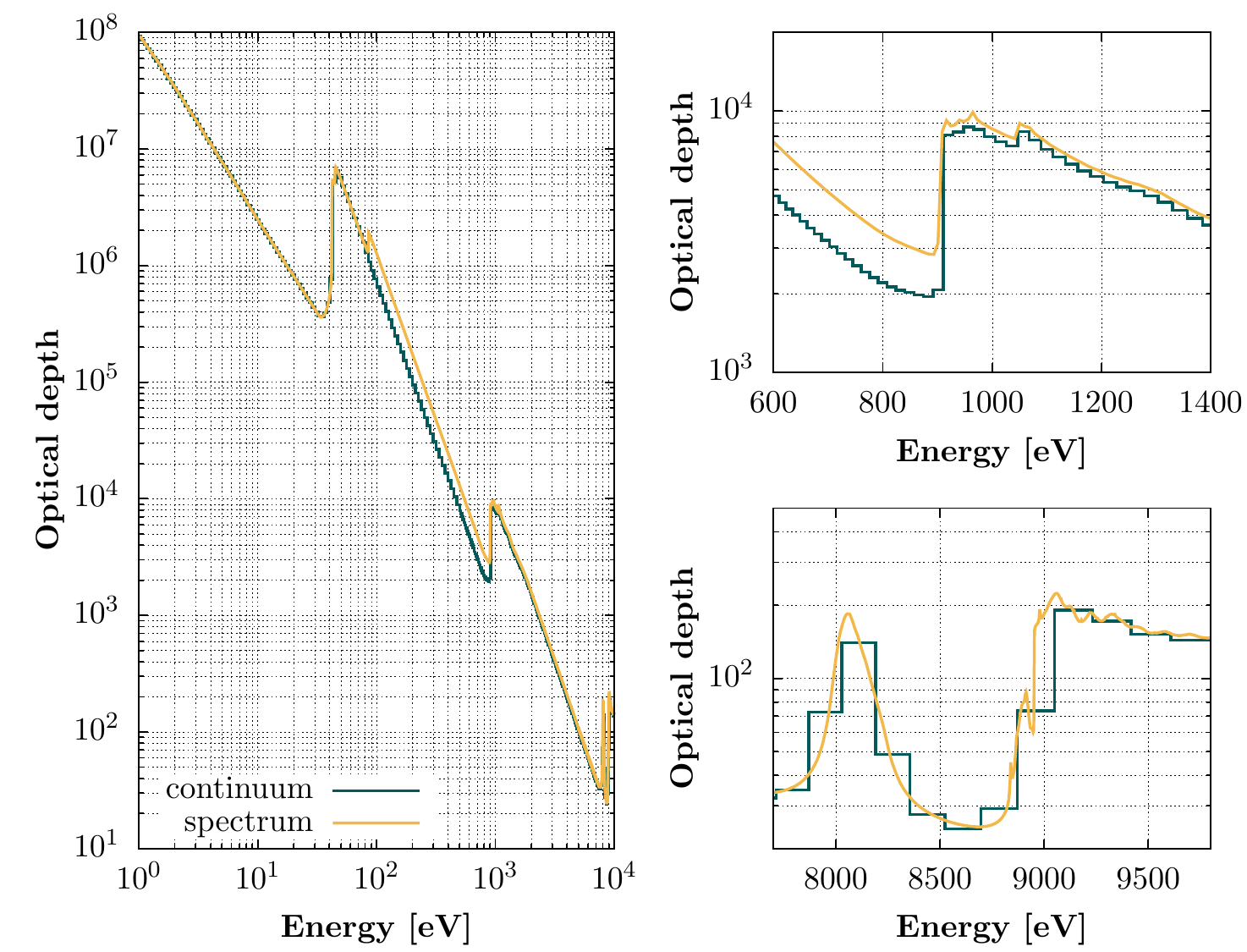}
    \caption{\label{fig:optical_depth}Comparing the time-integrated optical depth after 60 femtosecond simulation for a \(7\)~fs-long pulse with fluence $7{\times}10^5$~J/cm$^2$. The figure shows the continuum with 300 logarithmically-spaced bins. We used a high-resolution spectrum to check coverage of the continuum.}
\end{figure}

\newpage
\subsection{Converging the atomic model}
\begin{figure}[ht]
    \includegraphics[width=\linewidth]{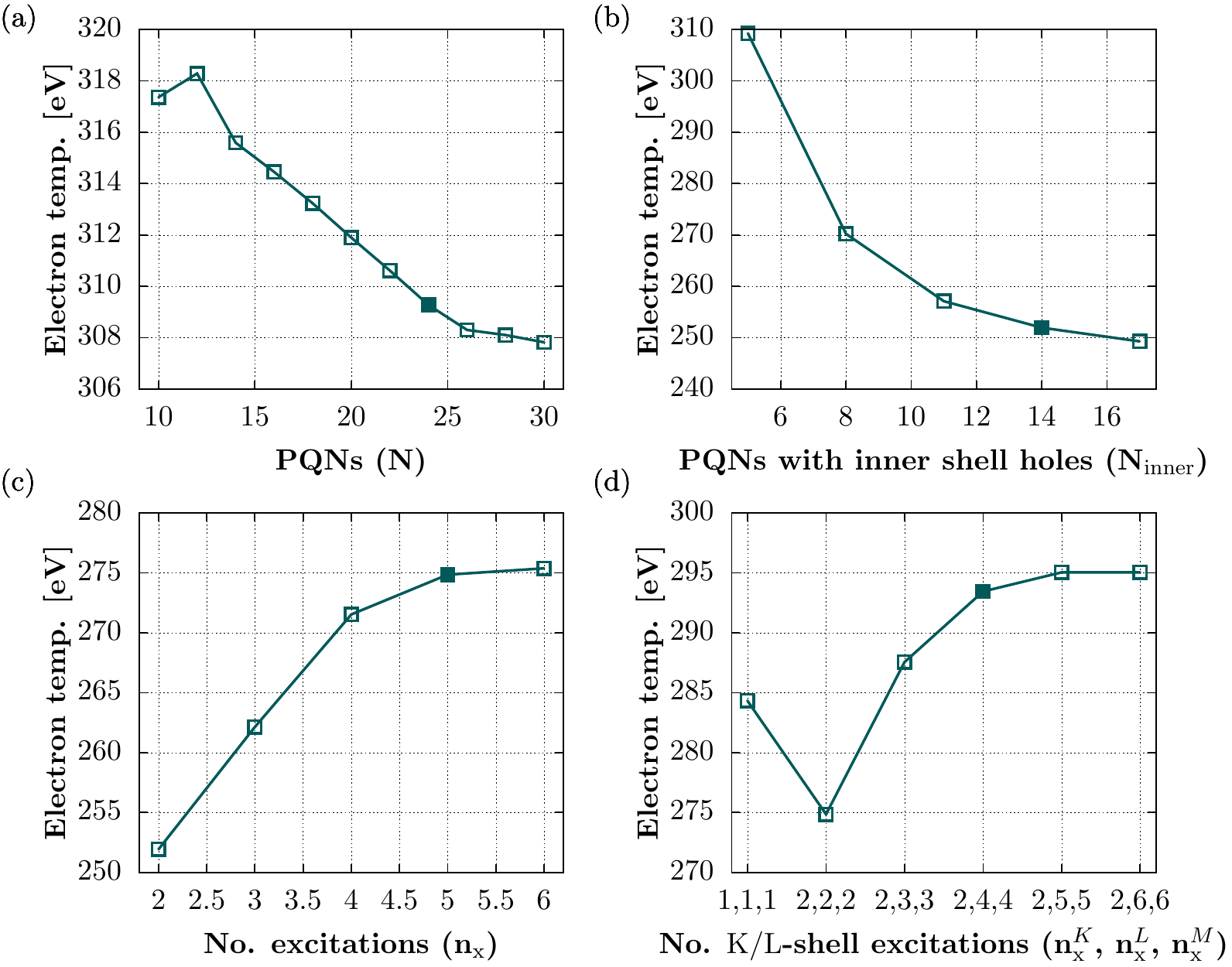}
    \caption{\label{fig:shm_convergence}SHM convergence results based on \(3.5{\times}10^{5}\)~J/cm\(^2\) incident fluence. We expected large PQNs values to contribute negligibly to the description of copper due to continuum lowering, whose effect increases with free electron density and decreases with increasing temperature. During the simulations, free electrons evolved to the extent where PQN states reappeared and became relevant to the properties of the system.}
\end{figure}

The task is to define a sufficiently inclusive model that contains all the important transition channels and population levels. We can reach an inclusive model by adding complexity until we see only moderate changes in the material's energy density. This is because spectral features that either influence the radiation distribution or materials properties, are already present. We achieved fully converged spectra by increasing the number of atomic levels while monitoring changes in the free electron temperature. One at a time, we adjusted parameters controlling the construction of the SHM atomic model, identified converged values from results in figure~\ref{fig:shm_convergence}, and incorporated these to determine successive parameters. We began by selecting the number of PQNs N\(=24\), followed by the number of PQNs with inner shell holes N\(_\mathrm{inner}=14\), maximum number of core-level excitations n\(_\mathrm{x}=5\), and finally the maximum number of \(\mathrm{K}\)/\(\mathrm{L}\)/\(\mathrm{M}\)-shell excitations n\(_\mathrm{x}^{K}\), n\(_\mathrm{x}^{L}\), n\(_\mathrm{x}^{M}=2, 4, 4\). Selected values reflect a trade-off between inclusiveness and computational demand.

\newpage
\subsection{Degeneracy and Coulomb coupling strength}
\begin{figure}[ht]
    \centering
    \includegraphics[width=0.75\linewidth]{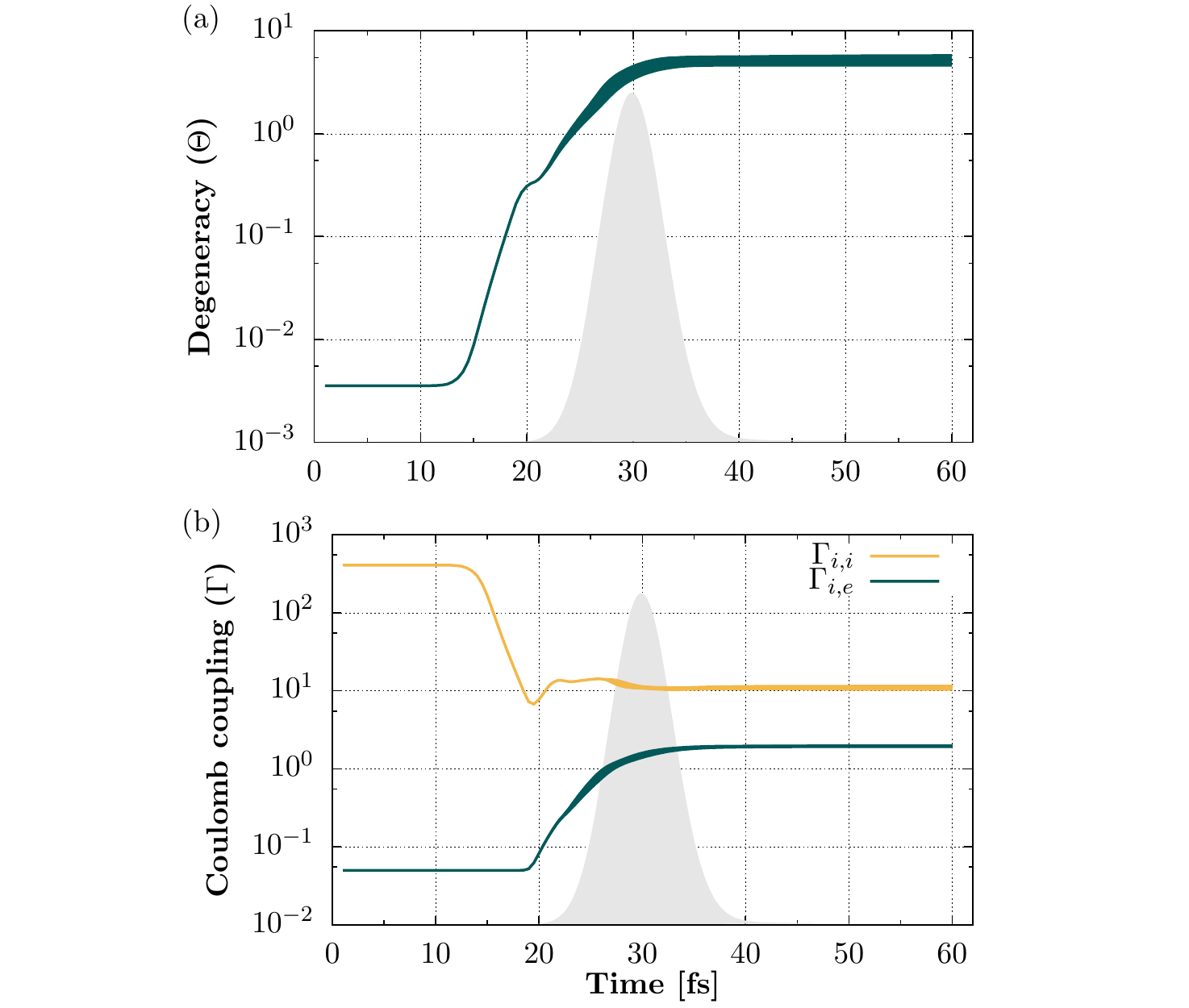}
    \caption{\label{fig:coupling}(a) Electron degeneracy and (b) Coulomb coupling parameter for ions and electrons for an incident fluence of \(3.5{\times}10^{5}\)~J/cm\(^2\). The thickness of the lines represents the standard deviations of these properties over the thickness of the sample. The beam profile is drawn in the background.}
\end{figure}

The degeneracy and Coulomb coupling parameters in general describe the state of a plasma. The degeneracy parameter \(\Theta\equiv T_{e}/T_{F}\), where \(T_{e}\) is the electron temperature and \(T_{F}\) is the Fermi temperature, determines the condition for a classical or quantum plasma depending on distance between electrons. The Coulomb coupling strength for ions \(\Gamma_{i,i}\equiv\langle U\rangle/\langle K\rangle\), where \(\langle U\rangle\) is the average internal energy and \(\langle K\rangle\) is the average kinetic energy per particle, evaluates the effects from the electrostatic forces~\cite{rightley_kinetic_2021}. In the case coupling between ions and electrons, we evaluated \(\Gamma_{i,e}\propto\Gamma_{i,i}\thinspace\Theta^{-3/2}/Z\thinspace\Theta^{-5/2}\) with \(Z=29\) which is valid in the completely degenerate limit where \(e^{\mu_{e}/T_{e}} >> 1\) and the chemical potential of the system \(\mu_{e}\rightarrow T_{F}\)~\cite{rightley_kinetic_2021,melrose_dispersion_2010}. We computed these three quantities over a single incident fluence with results presented in figure~\ref{fig:coupling}. The large Coulomb coupling strength \(\Gamma_{i,i}>>1\) at time zero implies ions in the plasma were strongly coupled indicative of a solid where many-body interactions dictate particle dynamics. As the energy was pumped into the system, we saw a solid to liquid transition reaching \(\Gamma_{i,i}\approx 10\), where kinetic and potential energies were comparable. At the beginning of the simulation, due to the low temperatures and occupation of the conduction band, free electrons were degenerate (\(\Theta<<1\)) and weakly coupled to the ions (\(\Gamma_{i,e}<1\)). Ionization pushed free electrons into a less degenerate but coupled regime. Together these results show copper reached the threshold between condensed and warm dense matter well before the end of the pulse profile.

\newpage
\bibliographystyle{unsrt}
\bibliography{ms}